\documentclass[11pt]{article}
\usepackage{fleqn,y10_style}
\usepackage{url}

\usepackage{graphicx}
\usepackage{epsfig}

\usepackage[figuresright]{rotating}

\newcommand{\beq}{\begin{equation}}
\newcommand{\eeq}{\end{equation}}

\newcommand{\apj}{{\em ApJ}}
\title{LOOPTOP AND FOOTPOINT IMPULSIVE HARD X-RAYS AND STOCHASTIC ELECTRON 
ACCELERATION IN FLARES}

\author{Vahe' Petrosian\address{Center for Space Science and Astrophysics,
        Stanford University, Stanford, CA 94305, USA}}

\begin{document}

% Typeset the front matter.
\maketitle

\begin{abstract} 

The discovery of hard X-rays from tops of flaring loops by the HXT of YOHKOH
represents a significant progress in the understanding of solar flares.  This
report describes the properties of 20 limb flares observed by YOHKOH from
October 1991 to August 1998, 15 of which show detectable impulsive looptop
emission.  Considering the finite dynamic range (about a decade) of the
detection it can be concluded that looptop emission is a common feature of all
flares.  The light curves and images of a representative flare are presented 
and
the statistical properties of the footpoint and looptop fluxes and spectral
indexes are summarized.  The importance of these observations, and those
expected from HESSI with its superior angular, spectral and temporal 
resolution,
in constraining the acceleration models and parameters is discussed briefly.

\end{abstract}

\section*{INTRODUCTION}

The most significant discovery of the HXT instrument on board the YOHKOH
satellite has been the detection of hard X-ray emission from the top of solar
flare loops as well as their footpoints.  The first so-called ``Masuda'' flare
is that of January 13, 1992 (Masuda et al.  1994; see also Alexander \& 
Metcalf
1997), which is clearly delineated by a soft X-ray (thermal) loop, and shows
three compact hard X-ray sources, two located at the footpoints (FPs) and a
third near the loop top (LT).  Several other such sources are described in
Masuda's thesis (1994).  As pointed out by Masuda et al.  (1994), these
observations lend support to theories that place the location of flare energy
release high up in the corona.  The power law hard X-ray spectra of the LT
sources indicate that electron acceleration is indeed occurring at or near 
these
locations.  The exact mechanism of the acceleration is a matter of 
considerable
debate.  In previous works (see Petrosian 1994 and 1996) we have argued that
among the three proposed particle acceleration mechanisms (electric fields,
shocks, and plasma turbulence or waves) the stochastic acceleration of ambient
plasma particles by plasma waves provides the most natural mechanism and can
explain the observed spectral features of flares (Park, Petrosian \& Schwartz
1997; hereafter {\bf PPS}).  In two recent works (Petrosian \& Donaghy, 1999 
and
2000; {\bf PD}) we demonstrated that the observed characteristics of the 
Masuda
flares can be used to constrain the model parameters.  In order to gain a
clearer picture of the frequency of occurrence of LT sources and the relative
values of the fluxes and spectral indexes of the FP and LT sources, we
(Petrosian, Donaghy \& McTiernan 2002; {\bf PDM}) have expanded and extended
Masuda's analysis.  In the next two sections I first summarize the results of
this work and then comment on their consequence for the acceleration 
mechanism.

\section*{DATA ANALYSIS AND RESULTS}

We have used The YOHKOH HXT Image Catalogue (Sato et al.  1988) to search for
flare candidates for detection of LT emission.  We have used Masuda's (1994)
selection criteria (heliocentric longitude $>80$ degrees, peak count rate $>10$
counts per sec per subcollimator in the $\sim 33-53$ keV range, {\it i.e.}  the
M2 channel).  We found 20 such events from 10/91 through 8/98, of which 11 were
selected by Masuda for the period of 10/91 to 9/93.  Observations of two events
are interrupted by spacecraft night.  Of the remaining 18 events, 15 show
detectable impulsive looptop emission.  As described below, considering that the
finite dynamic range (about a decade) of the detection introduces a strong bias
against observing comparatively weak looptop sources, one can conclude that LT
emission is a common feature of all flares.  An interesting aside, is that of
the 9 new events, 3 appear to be examples of interacting loop structures with
multiple LT and FP sources, of the type analyzed by Aschwanden et al.  (1999).
It is surprising that none of the 11 Masuda events are in this category.

\begin{figure}[htb]
\epsfig{file=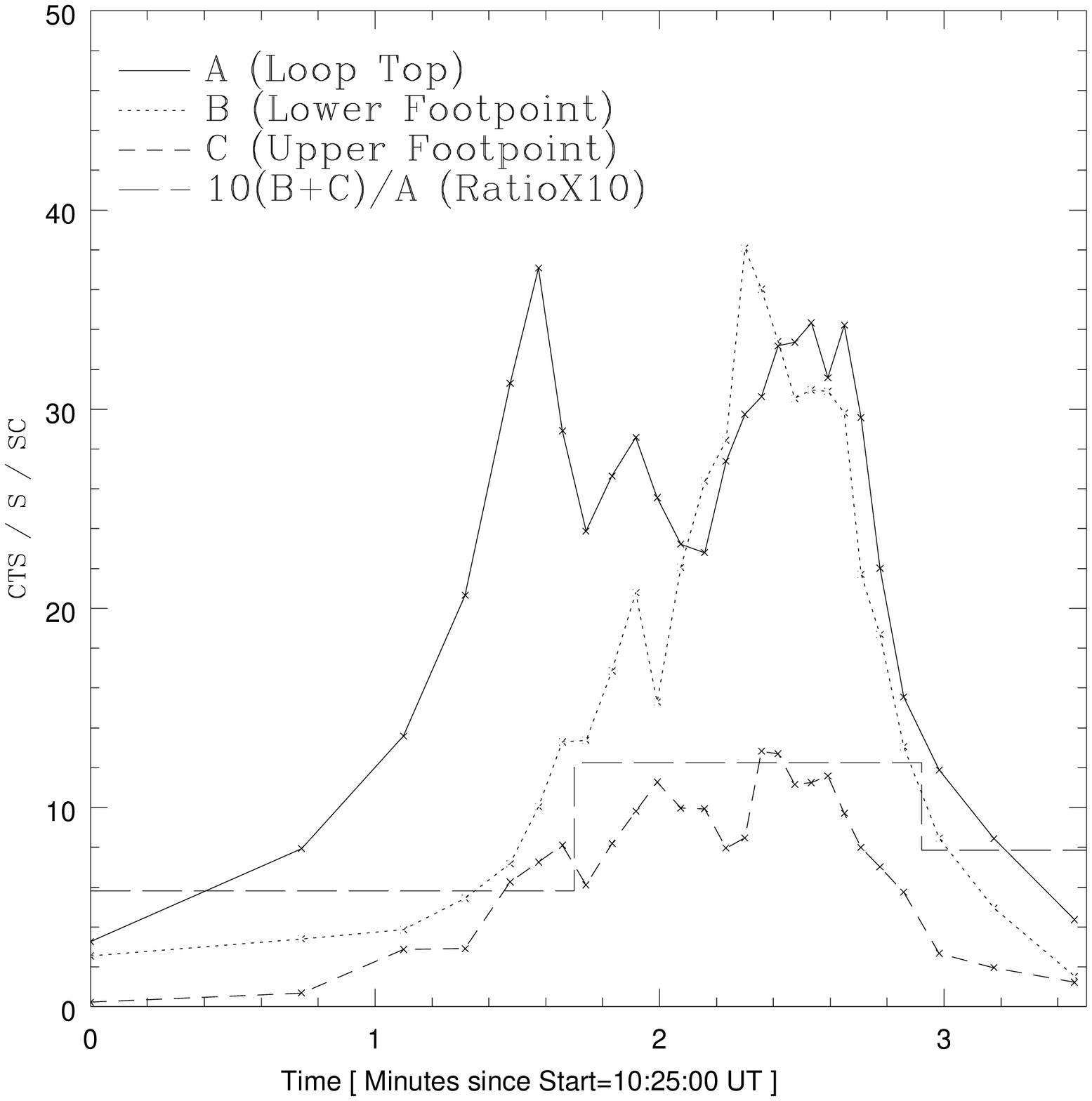,width=3.3in,height=3.3in}
\psfig{file=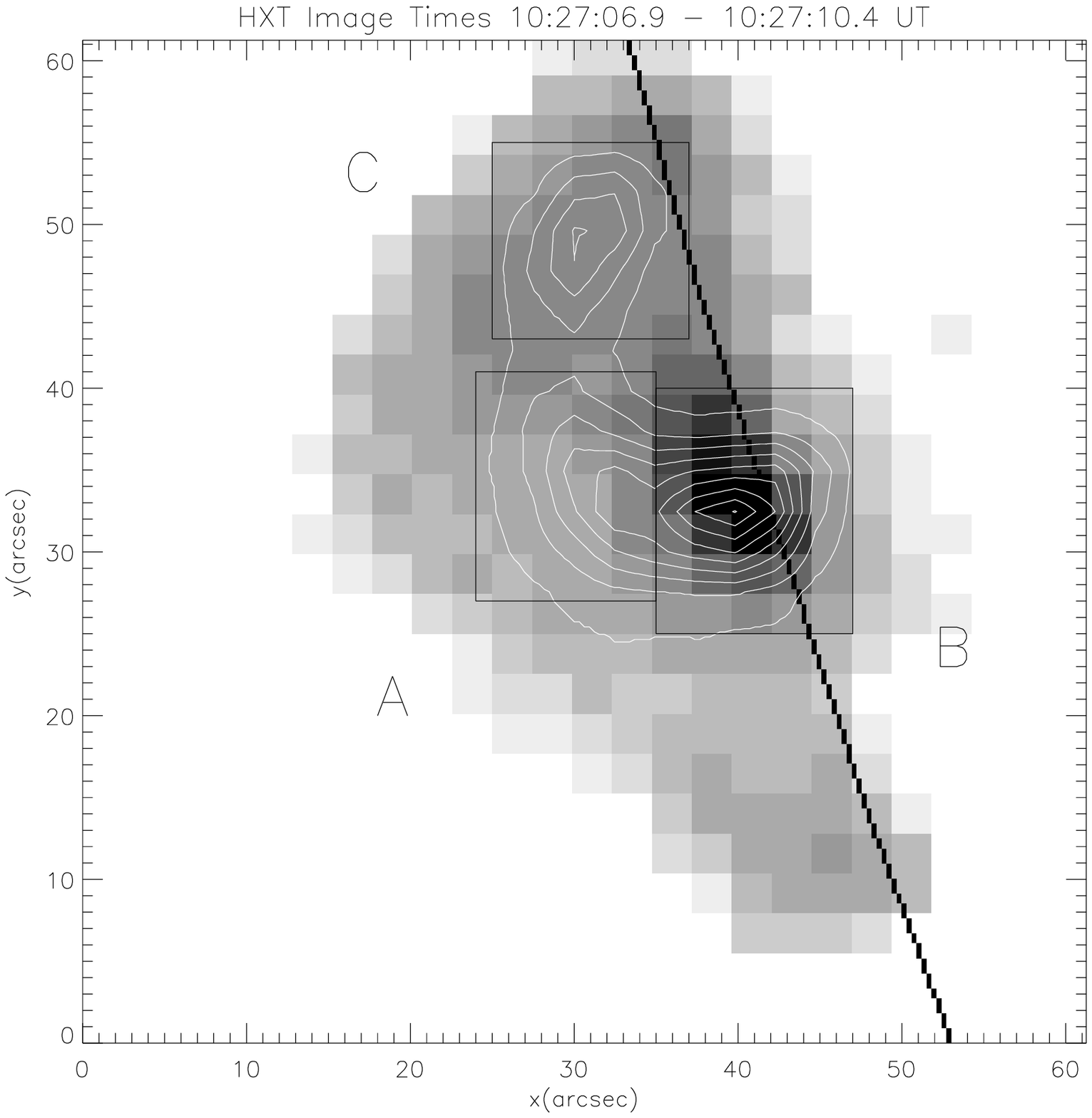,width=3.1in,height=4.2in,angle=90} 
\caption{Images ({\bf right panel}) and light curves ({\bf left panel}) for the
December 18, 1991 flare.  The contours and the gray scale show the HXT (channel
M2; $\sim 33-53$ keV) and SXT images of the loop, respectively, for the
specified time.  The diagonal line shows the location of the solar limb.  The
brightest contour and the contour separations are $B_{max}=8.1$ and $\Delta
B=0.73$ counts/pixel with 2.5 sq.  arc second size pixels.  The light curves of
the the LT and FP sources refer to the counts integrated over regions shown on
right panel.  The dashed histogram shows the average of the ratio of counts 
${\cal R}=FPs/LT$ (multiplied by 10) for three time intervals.}
\end{figure}

We have constructed HXT images and investigated their evolution throughout all
these flares using the YOHKOH spectral and spatial analyses software packages.
In a few cases we have also used the Alexander \& Metcalf (1997) ``pixon" method
of image reconstruction.  From the investigations of these images we have
determined the locations of LT and FP sources and produced separate light curves
for the well defined sources.  Figure 1 shows an example of a simple loop with
an intermediate strength LT source and Figure 2 shows a flare with a more
complex morphology consisting of two loops with different but related temporal
evolution.

\begin{figure}[htb]
\epsfig{file=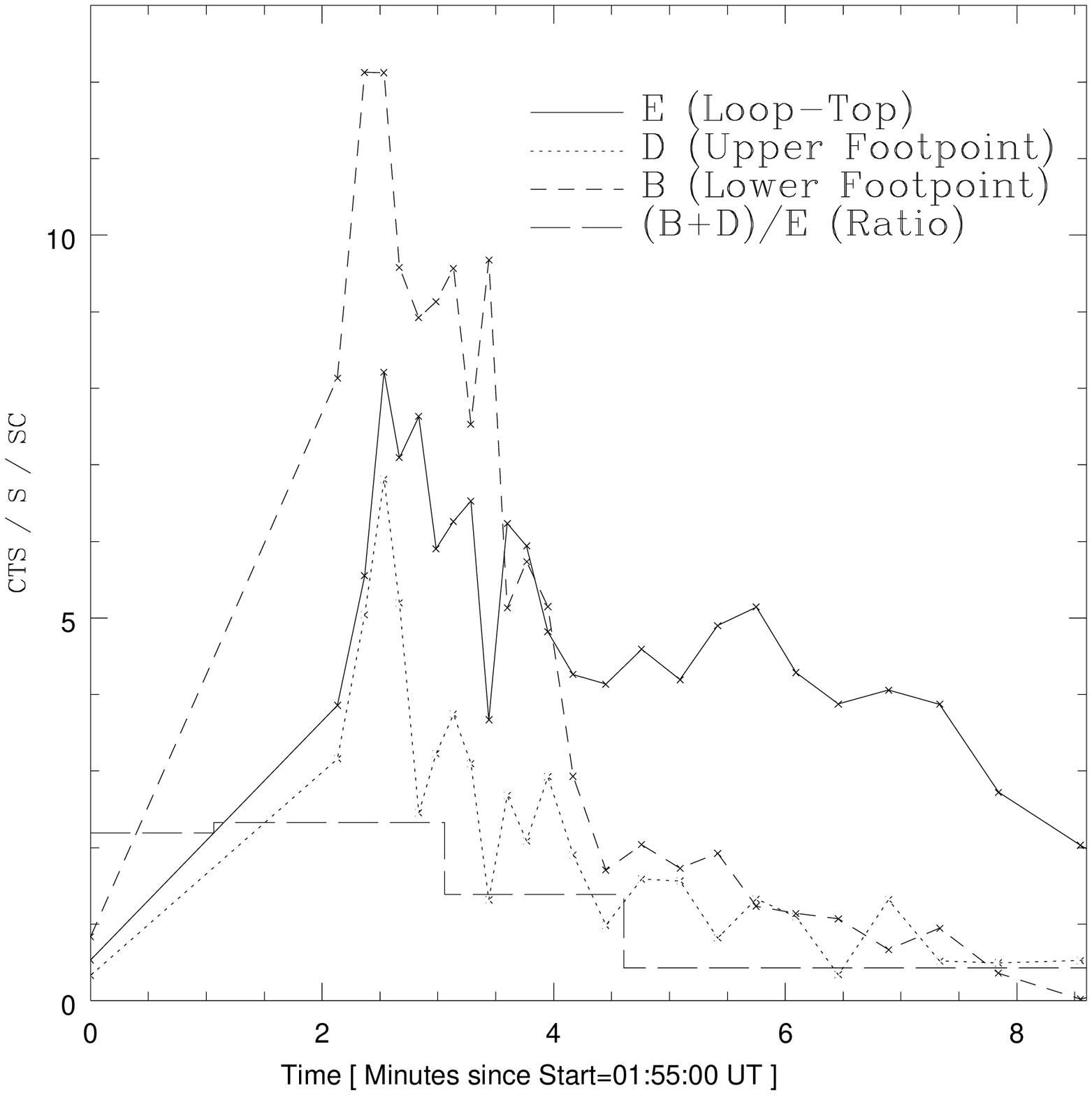,width=3in,height=2.75in}
\epsfig{file=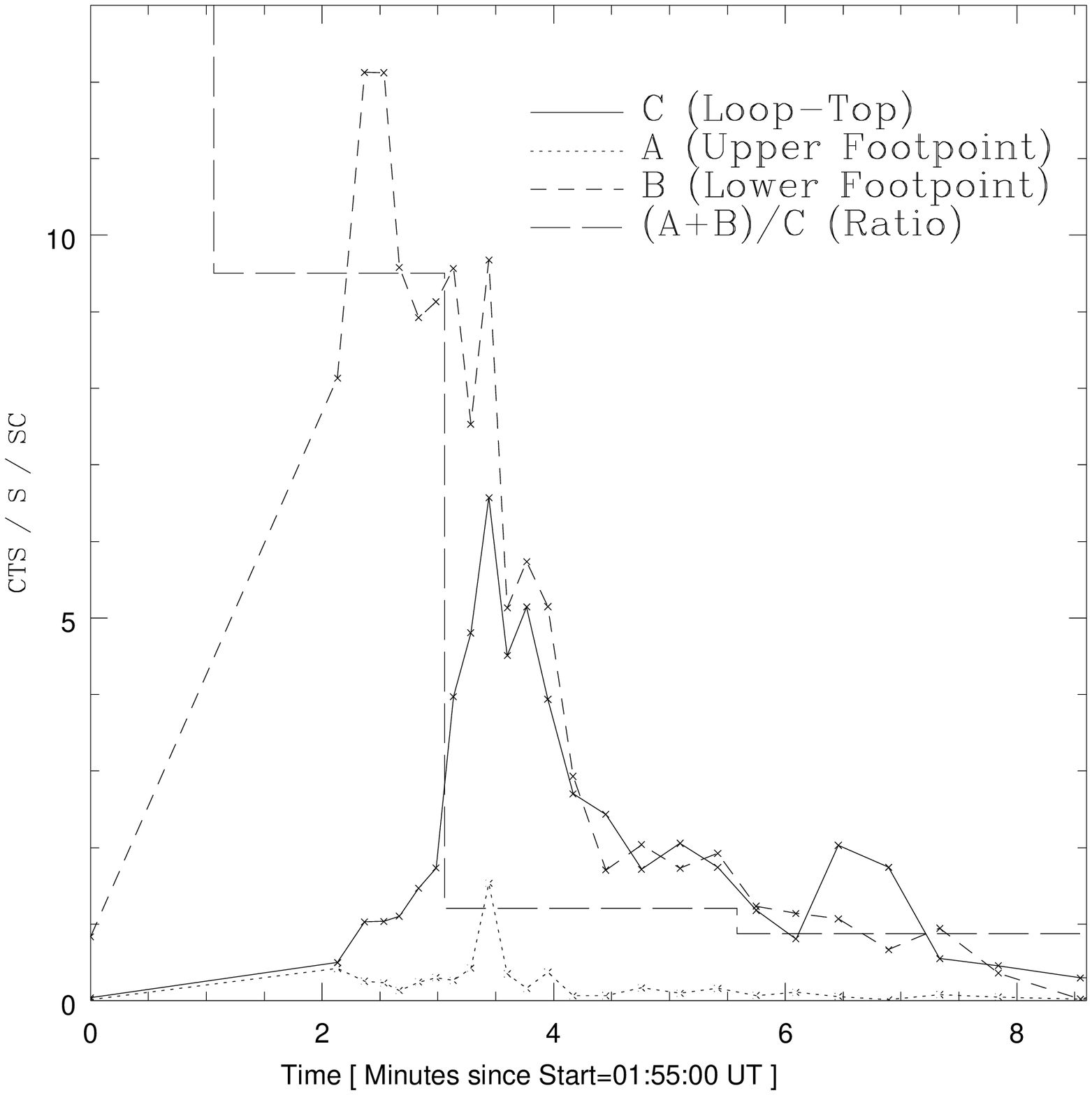,width=3in,height=2.75in}
\psfig{file=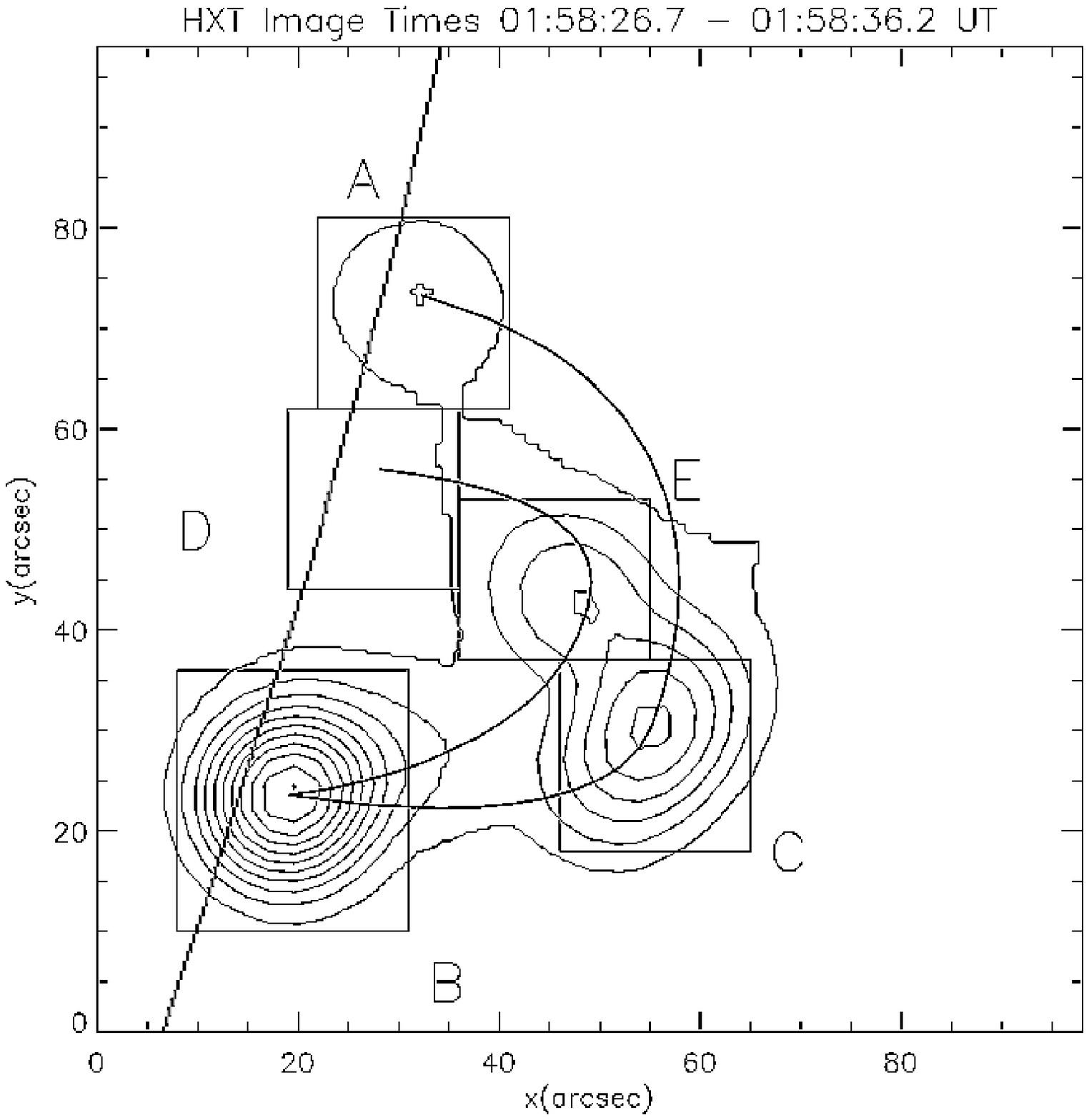,width=3.85in,height=2.75in}\hspace{0.4cm}
\psfig{file=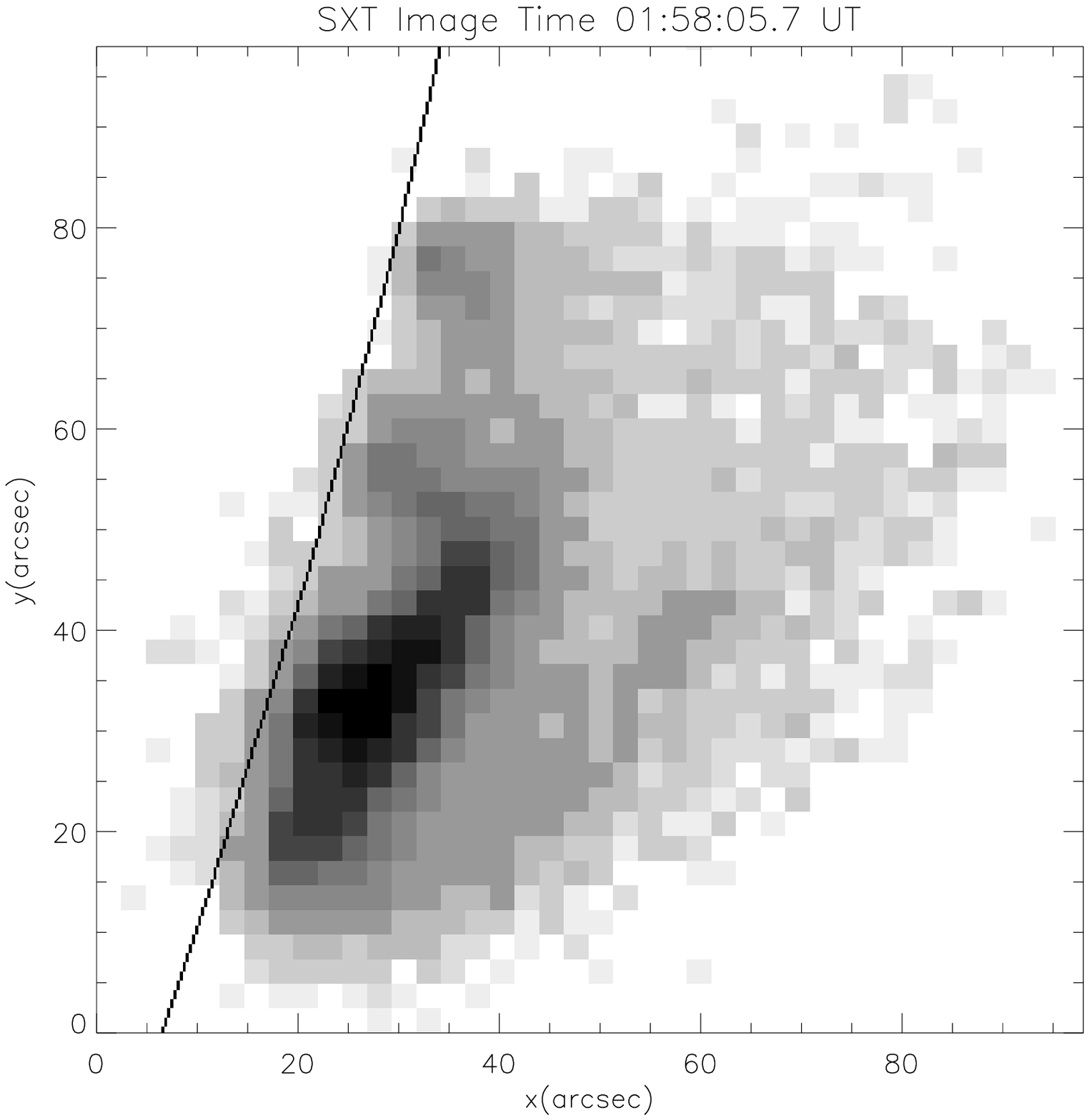,width=2.75in,height=3.8in,angle=90}
\caption{Same as Figure 1 for the August 18, 1998 flare.  The upper left and
right panel light curves represent the southern (AEB) and the northern (BDE)
loops, respectively.  Note that for the LT source D we plot counts divided by 3.
In the HXT image (lower left panel) $B_{max}=14.8$, $\Delta B=0.82$
counts/pixel, the digonal line shows the limb location, and the two arcs sketch
the presumed loop outlines. The SXT image shown on the lower
right panel was taken nearly two minutes after the HXT image.}
\end{figure}

We determine the relative fluxes of the LT and FP sources and obtain rough
measures of some of the spectral characteristics ({\it e.g.}  power-law
indexes).  Figure 3 shows the M1 channel ($\sim 23-33$ keV) counts of the FPs vs
LT sources for all flares (left panel) and the distributions of the count ratio
${\cal R}=FPs/LT$ (right panel).  We use a representative time period around an
impulsive peak and avoid the later stages (the third periods of the histograms
shown along the light curves) which can be contaminated by thermal emission.
Note that some flares (those connected by dashed lines contribute more than one
data point.

\begin{figure}
\epsfig{file=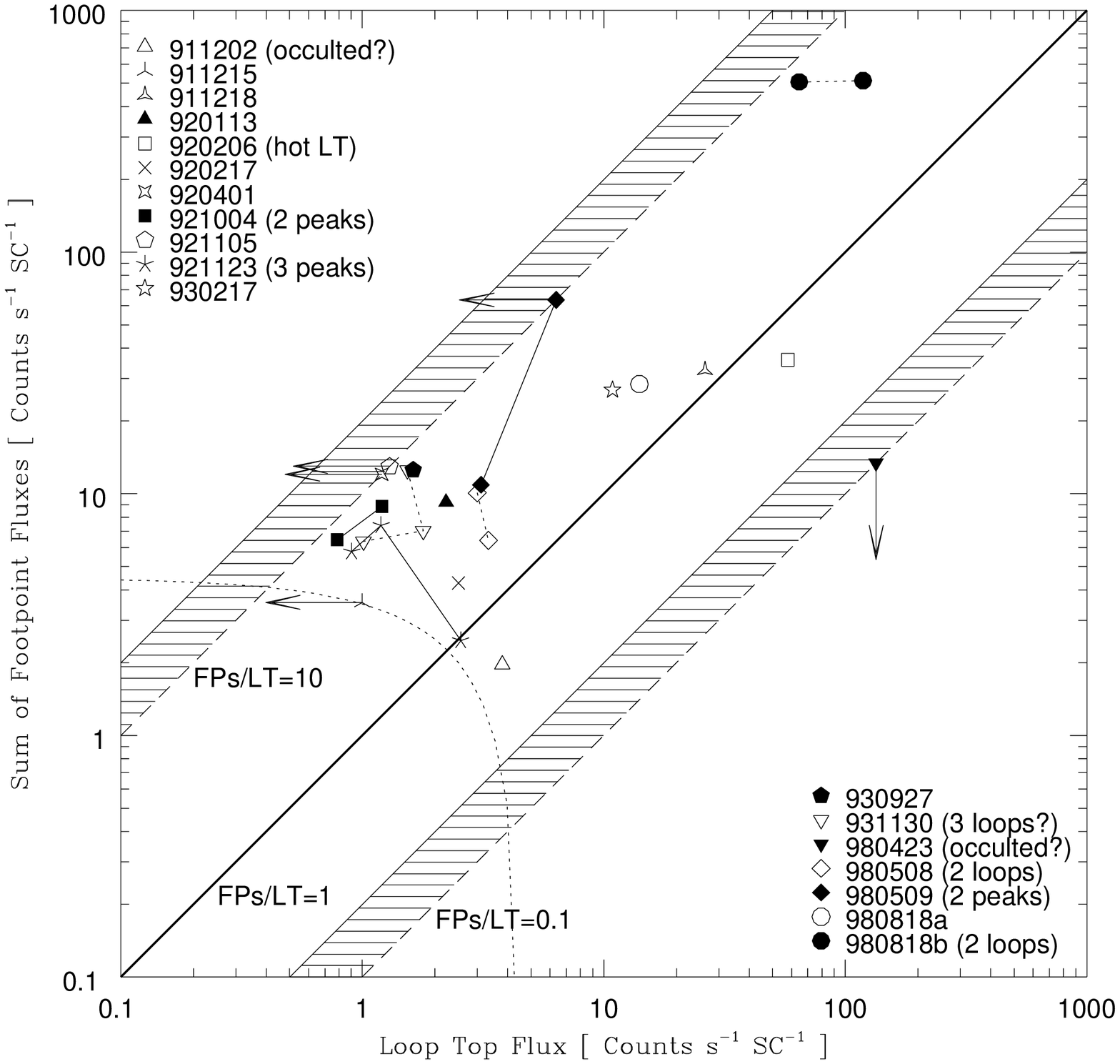,height=3.1in}\hspace{1cm}
\epsfig{file=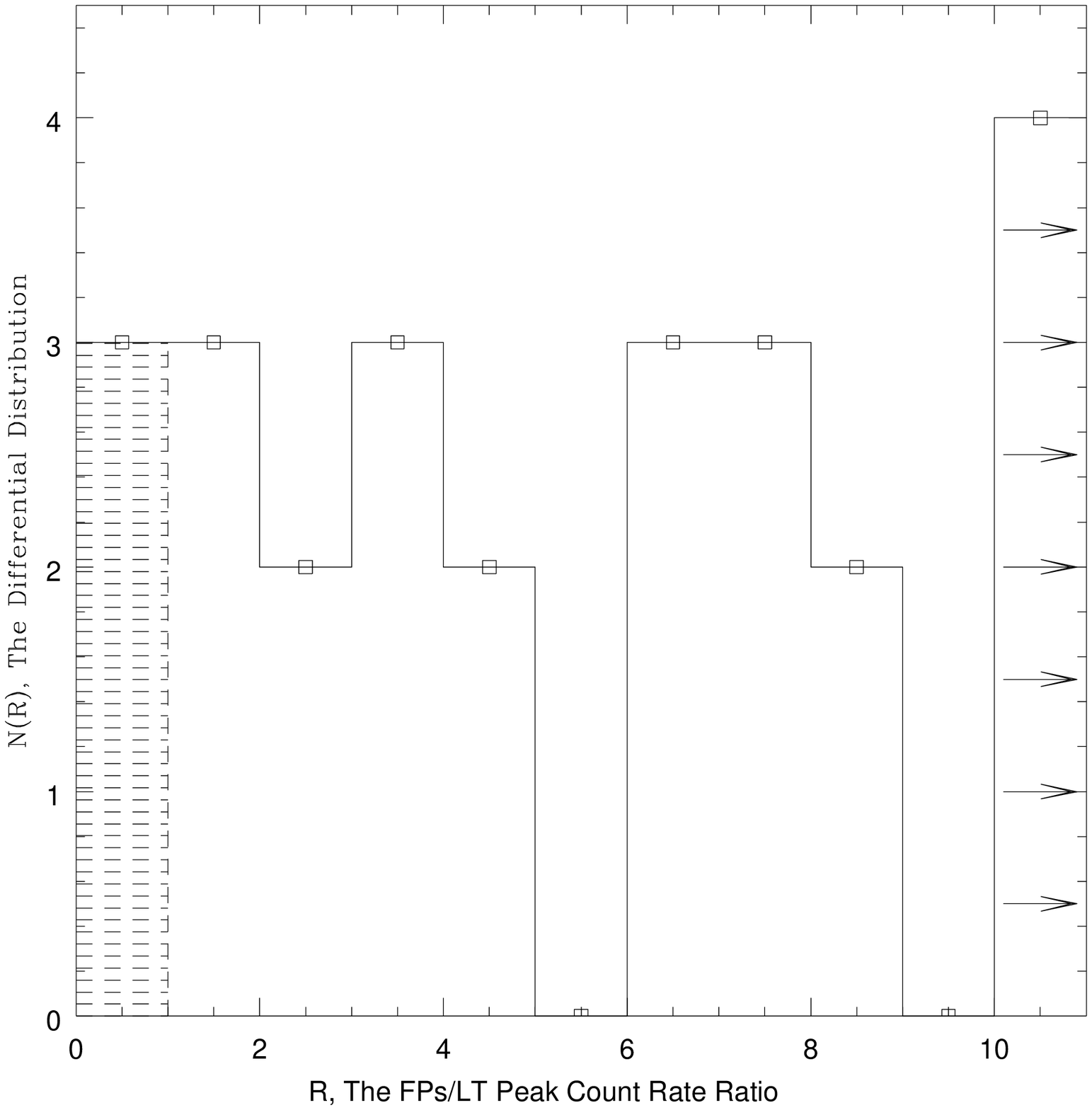,height=3.1in}
\caption{({\bf Left Panel}):  Counts from two FPs vs LT counts, in the M1
channel.  The diagonal lines show lines of constant ratio (${\cal R}=FPs/LT$)
and represent detection thresholds arising from the finite dynamic range of the
instrument which is about 10.  Flares with undetected LT source are denoted by
an arrow placed on the upper bound of detection of ${\cal R}=10$.  The dotted
curve shows the event selection threshold of 10 counts at the M2 channel.  ({\bf
Right panel}):  The differential distribution of the ratio ${\cal R}=FPs/LT$ of
all flares.  The arrows indicate ratios greater than the dynamic range.  Some of
the flares in the shaded area with ${\cal R}<1$ may be occulted or be dominated
by a superhot thermal component.}
\end{figure}

Analysis of these results lead to  the following very important conclusions (see 
also {\bf PDM}). 

$\bullet$ The LT hard X-ray emission seems to be a common characteristic of the
impulsive phase of solar flares, appearing in some form in 15 of the 18 selected
flares.  The absence of LT emission in the remaining cases (those indicated by
the horizontal arrows in Figure 3) is most likely due to the finite dynamic
range of the imaging technique which is about 10.  The hatched diagonal regions
show this range.  Flares outside the area between these two bounds will have
either a too weak a FP or LT source to be detected by HXT.  From this we
conclude that {\em LT emission is present in all flares.}  However there are
very few flares with ${\cal R} < 1$ and there are indications that the three
such cases seen in Figure 3 are either partially occulted or are dominated by a
superhot LT source.  Thus one may conclude that, in general, the ratio ${\cal
R}$ has a relatively flat distribution between 1 and 10, with few cases outside
this range.  A larger sample with a wider dynamic range will be required for a
better determination of this distribution.

Figure 4 shows the distribution of the overall spectral index (left panel) and
the distribution of the difference between low ($\sim 13-28$ keV, L and M1
channels) and high ($\sim 28-53$ keV, M1 and M2 channels) energy indexes.
Clearly with only a four channel data one must be cautious in the interpretation
of these histograms.  Nevertheless, some significant conclusions can be drawn
from these results as well.

$\bullet$ The overall distribution of the power-law spectral index $\gamma$
rises rapidly above 2, peaks around 4 and then declines gradually thereafter.
This is similar to previous determinations of this distributions from HXRBS on
board the {\it Solar Maximum Mission} (see {\it e.g.}  McTiernan \& Petrosian
1991), but contains a few more steep spectra, specially for LT sources.  This
difference could be due to thermal contamination and/or because HXT is sensitive
to lower photon energies than HXRBS.  On the average, the spectral index of LT
sources is larger ({\it i.e.}  spectra are steeper) than that of the FP sources
by one unit; ${\bar\gamma}_{LT}=6.2\pm 1.5, {\bar\gamma}_{FP}=4.9\pm 1.5$.  The
physics of the acceleration process must certainly play a role here.

$\bullet$ The spectra tend to steepen at higher energies (spectral index
$\gamma$ increases by 1 to 2), especially for sources with $\gamma <6$, for
which the thermal contribution should be the lowest.  This is the opposite of
what is observed at higher energies, where spectra tend to flatten above 100's
of keV (McTiernan \& Petrosian 1991).  The directivity of the X-ray emission and
the albedo effect for the limb flares under consideration could play some role
here, especially for the FP sources.

$\bullet$ Finally, we note that solar flares occur in many different
morphologies, the most common being a simple flaring loop with one LT and two FP
sources.  However, interacting loop models and even more complicated structures
are frequently observed.  There is a hint that the frequency of occurrence of
complex morphologies may be different for the declining and growing phases of
the solar cycle.

\begin{figure}
\epsfig{file=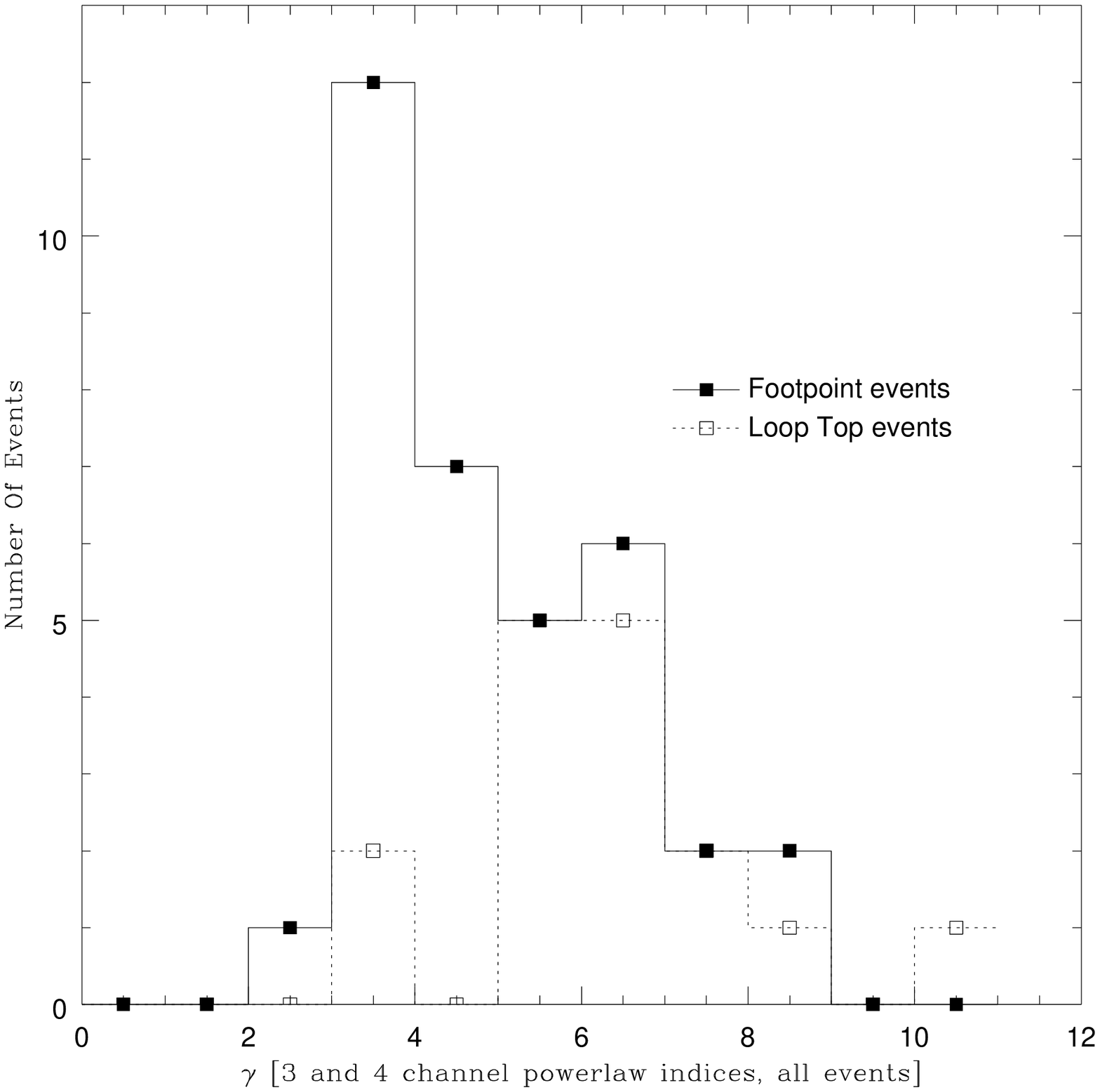,height=3.3in}
\epsfig{file=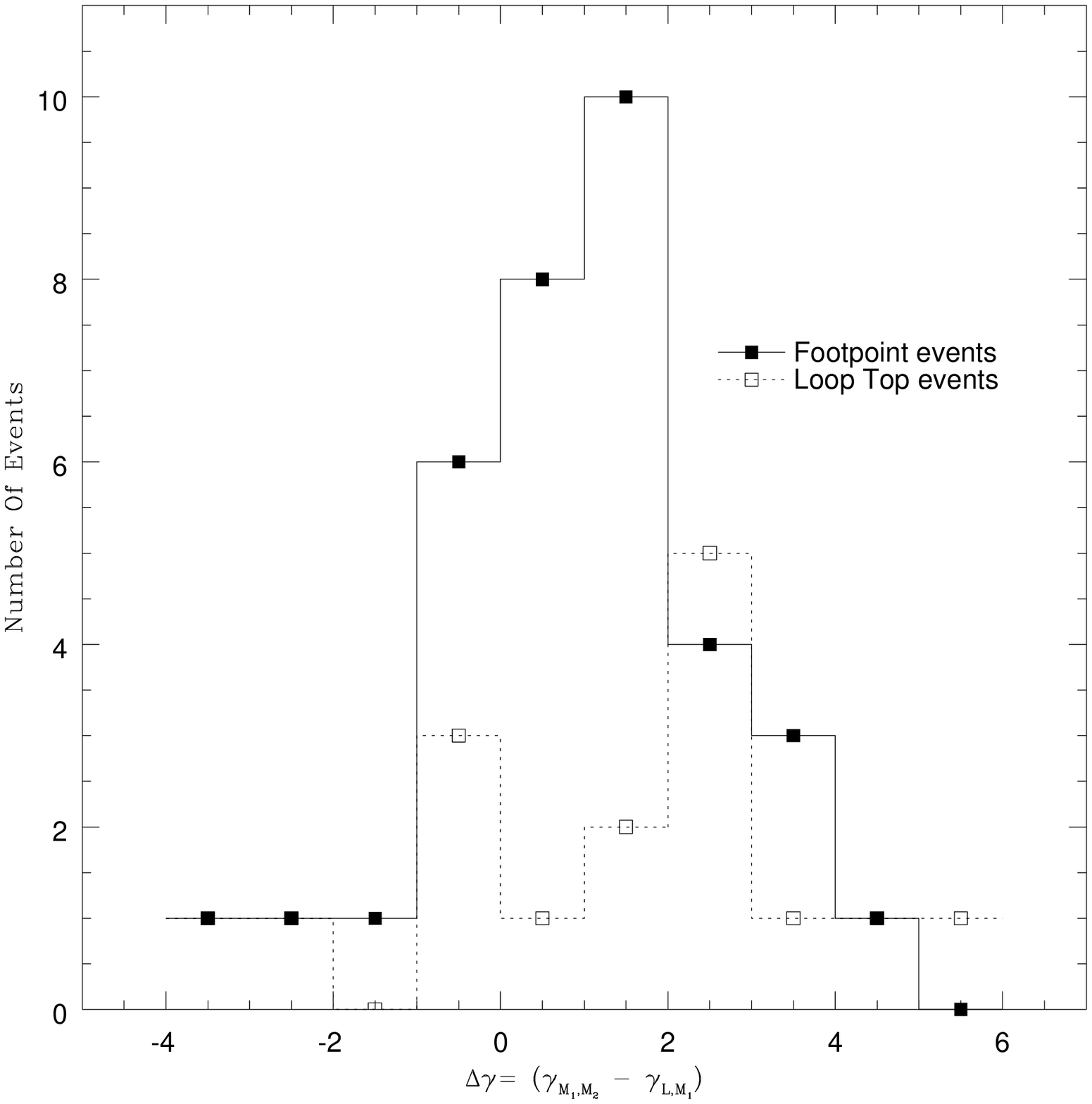,height=3.3in}
\caption{ The distribution of the overall spectral power-law index $\gamma$
({\bf Left Panel}), and the distribution of
$\Delta\gamma=\gamma_{M1,M2}-\gamma_{L,M1}$ ({\bf right panel}).  The solid
histograms and filled points represent the FP sources and the dotted histogram
and open points represent the LT sources.}  
\end{figure}

\section*{THEORETICAL IMPLICATIONS}

The above results can be used to constrain the model parameters describing the
plasma in the acceleration site and those describing the acceleration mechanism.
These parameters define several important timescales:  The acceleration time
scale is related to the energy diffusion coefficient $D_{EE}$ as $\tau_{\rm ac}
\sim E^2/D_{EE}$.  The mean scattering time is inversely proportional to the
pitch angle diffusion coefficient, $\tau_{\rm sc} \sim 1/D_{\mu \mu}$.  The time
for a particle with velocity $v$ to cross the acceleration site of size $L$ is
$\tau_{\rm tr} \sim L/v$; these two timescales determine the escape time from
the acceleration site (for $\tau_{\rm sc}< \tau_{\rm tr}, \, T_{\rm esc} \sim
\tau_{\rm tr}^2/\tau_{\rm sc}$, otherwise $T_{\rm esc} \sim \tau_{\rm tr}$).
Finally the energy loss timescale for an electron of energy $E$ is $\tau_{\rm
L}=E/{\dot E}_{\rm L}$, which for the non relativistic electrons under
consideration here is dominated by the Coulomb losses, $\tau_{\rm Coul}= vE/(4
\pi r_{0}^{2} {\rm ln} \Lambda n mc^4)$, where $4 \pi r_{0}^{2} {\rm ln}
\Lambda= 2 \times 10^{-23}$ cm$^2$ and $m$ is the mass of the electron.  The
values of these time scales depend on the plasma density $n$, magnetic field
$B$, plasma turbulence energy density $w_{\rm turb}$ and size $L$, and their
variations with energy depend on these parameters and the spectrum of the
turbulence (for details see {\bf PPS} and {\bf PD} and reference cited there).

For example, if $T_{\rm esc}$ is large the accelerated electron spend a long
time in the acceleration site or at the loop top giving rise to a strong LT
source.  Inversely, a weak LT source is expected for a short $\tau_{\rm Coul}$.
Very roughly, the ratio of the FP to LT emission is expected to vary as ${\cal
R} = J_{FPs}/J_{LT} \sim \tau_{\rm Coul}/T_{\rm esc}$, where the $J$'s refer to
the expected bremsstrahlung fluxes.  Furthermore, the spectral shape of these
fluxes are also related to the above mentioned parameters.  It is clear, for
example, that if the acceleration time is short compared to the escape time,
then more electrons get to higher energies resulting in a flat accelerated
electron spectrum and LT hard X-rays.  For a power law accelerated spectrum,
$f(E)\propto E^{-\delta}$, the LT ({\it thin target}) hard X-ray spectrum at
photon energy $k$ is $J_{LT}\propto k^{-\delta -1/2}$.  On the other hand, the
spectrum of the electrons that escape the acceleration region and reach the
footpoints is $f(E)/T_{\rm esc}(E)\propto E^{-\delta -s'}$, assuming that for
the small energy range of the HXT we can use the approximation $T_{\rm
esc}(E)\propto E^{s'}$.  These electrons will emit a {\it thick target} spectrum
at the footpoints with $J_{FPs}\propto k^{-\delta -s' + 1}$.  Thus, for LT
spectra that are steeper than the FPs spectra we require $s' < 3/2$.  When the
escape is determined by the traverse time $s'= -1/2$ and this condition is
satisfied.  And when scattering dominates, this requires $\tau_{\rm sc}\propto
E^s$, with $s > -5/2$.  The energy dependence of $\tau_{\rm sc}$ depends on the
characteristics of the turbulence.  In general, one expects a positive value for
$s$, and even for for a very steep spectrum of the turbulence one has $s> -1$
(see {\it e.g.}  Pryadko \& Petrosian 1997).

However, it should be noted that these relations are very approximate and valid
only for a limited energy range and very steep electron spectra; they breakdown
completely for $\delta < 2.5$.  Nevertheless, this excersize demonstrates that
using the observed values of the spectral indexes and FP and LT counts we can
determine the plasma and acceleration characteristics.  As shown in {\bf PD} the
values of the parameters such derived from the YOHKOH high spatial resolution
data are very reasonable, and agree with those derived by {\bf PPS} from fits to
large dynamic range overall spectra.  It is clear then that a more refined and
simultaneous observations of the flare characteristics can yield important
information about the the acceleration mechanism, the energy release and the
evolution of solar flares.  We eagerly anticipate the increased spectral,
temporal and spatial resolution possible with the instruments of the RHESSI
satellite.

\section*{ACKNOWLEDGEMENTS}

This work is supported in parts by NASA grants NAG-5-7144-0002 and  
NAG5-8600-0001. This paper was completed during the authors stay at the 
Institute for Theoretical Physics at UC Santa Barbara, which is supported in 
part by the National Science Foundation Under Grant No. PHY99-07949

\end{document}